\documentclass[12pt,letterpaper]{article}
\usepackage[usenames, dvipsnames]{xcolor}
\usepackage{jcapmod}

\usepackage{tocloft}
\usepackage[]{todonotes}
\usepackage{verbatim}
\usepackage[utf8]{inputenc}
\usepackage{mathrsfs}
\usepackage{cleveref}
\usepackage[shortlabels]{enumitem}

\usepackage{pgf}
\usepackage{tikz-cd}
\usetikzlibrary{shapes,arrows}
\usetikzlibrary{calc}
\usepackage{pgfplots}
\pgfplotsset{compat=1.12}
\usepackage{tikz-3dplot}

\usepackage{tikz}
\usepackage{setspace,caption}
\usepackage{lipsum}
\usetikzlibrary{matrix,arrows,calc}
\makeatletter
\newsavebox\myboxA
\newsavebox\myboxB
\newlength\mylenA

\definecolor{cornellRed}{HTML}{B31B1B}
\definecolor{cornellBlue}{HTML}{0068AC}
\definecolor{cornellGreen}{HTML}{6EB43F}

\newcommand{\FF}{\mathbb{F}}
\newcommand{\rset}{[\mathbf{r}]}

\newtheorem{theorem}{Theorem}
\newtheorem{lemma}[theorem]{Lemma}

\newtheorem{proposition}[theorem]{Proposition}

\usepackage{bbm} 					%%% Blackboard Bold math symbols
\usepackage{slashed} 				%%% Feynman slash
\usepackage{graphicx}				%%% Graphics
\usepackage{subcaption}			%%% Subcaptions for subfigures
\usepackage{psfrag}				%%% Editing EPS files in TeX
\usepackage{tensor}				%%% The \indices command
\usepackage{fouridx}				%%% Arbitrary subscripts/superscripts
\usepackage{bm}					%%% Bold Math symbols
\usepackage{mdframed}				%%% Box for \aside
\usepackage{multirow}				%%% More flexible tables
\usepackage{soul}					%%% Strike-through text
\usepackage{bbold}				%%% Blackboard Bold
\usepackage{multicol}				%%% Pages with multiple columns
\usepackage{tikz-cd}
\usepackage{rotating}

\usetikzlibrary{arrows}

\tikzset{
commutative diagrams/.cd,
arrow style=tikz,
diagrams={>=latex}}

\usepackage{amsmath}
\usepackage{amssymb}
\usepackage{amsfonts}
\usepackage{mathtools}

\usepackage{feynmf}
\usepackage{marvosym}

\usepackage{import}

\newcommand*\xoverline[2][0.75]{%
    \sbox{\myboxA}{$\m@th#2$}%
    \setbox\myboxB\null% Phantom box
    \ht\myboxB=\ht\myboxA%
    \dp\myboxB=\dp\myboxA%
    \wd\myboxB=#1\wd\myboxA% Scale phantom
    \sbox\myboxB{$\m@th\overline{\copy\myboxB}$}%  Overlined phantom
    \setlength\mylenA{\the\wd\myboxA}%   calc width diff
    \addtolength\mylenA{-\the\wd\myboxB}%
    \ifdim\wd\myboxB<\wd\myboxA%
       \rlap{\hskip 0.5\mylenA\usebox\myboxB}{\usebox\myboxA}%
    \else
        \hskip -0.5\mylenA\rlap{\usebox\myboxA}{\hskip 0.5\mylenA\usebox\myboxB}%
    \fi}
\makeatother

\newcommand{\cE}{\mathcal{E}}

\newcommand{\cK}{\mathcal{K}}
\newcommand{\cL}{\mathcal{L}}

\newcommand{\cO}{\mathcal{O}}

\newcommand{\cV}{\mathcal{V}}

\newcommand{\cB}{\mathcal{B}} % Mike added

\newcommand{\PP}{\mathbb{P}}

\newcommand{\ZZ}{\mathbb{Z}}

\definecolor{cobalt}{RGB}{44, 98, 120}
\definecolor{celadon}{rgb}{0.67, 0.88, 0.69}
\definecolor{dm}{cmyk}{.20, 0, .30, 0}
\definecolor{burgundy}{rgb}{0.5, 0.0, 0.13}
\definecolor{plotBlue}{RGB}{94, 130, 181}

\newcommand{\C}{\mathbb{C}}

\newcommand{\awgc}{c}

\hypersetup{
  colorlinks,
  citecolor=Violet,
  linkcolor=cobalt,
  urlcolor=Blue}

\DeclareSymbolFontAlphabet{\mathbb}{AMSb}

\newif\iffastcompile

\fastcompilefalse
\iffastcompile
\newcommand{\cl}[1]{}
\newcommand{\lm}[1]{}
\newcommand{\md}[1]{}
\newcommand{\ab}[1]{}
\else
\newcommand{\cl}[1]{\todo[color=burgundy!30, size=\scriptsize, bordercolor=burgundy!30]{CL: #1}}
\newcommand{\lm}[1]{\todo[color=dm!90, size=\scriptsize, bordercolor=dm!90]{LM: #1}}
\newcommand{\md}[1]{\todo[color=blue!30, size=\scriptsize, bordercolor=blue!50]{MD: #1}}
\newcommand{\ab}[1]{\todo[color=blue!30, size=\scriptsize, bordercolor=blue!50]{AB: #1}}

\fi

\newcommand{\email}[1]{\href{mailto:#1}{#1}}

\ProvideTextCommandDefault{\Dbar}{%
\leavevmode\lower.5ex\rlap{\hskip-.07em\accent"16}D%
}

\begin{document}
	\newcommand{\main}{.}
\begin{titlepage}

\setcounter{page}{1} \baselineskip=15.5pt \thispagestyle{empty}
\setcounter{tocdepth}{1}

\bigskip\

\vspace{1cm}
\begin{center}
{\fontsize{22}{28} \bfseries Minimal Surfaces and Weak Gravity}
\end{center}

\vspace{0.45cm}

\begin{center}
\scalebox{0.95}[0.95]{{\fontsize{14}{30}\selectfont  Mehmet Demirtas,$^{a}$ Cody Long,$^{b}$ Liam McAllister,$^{a}$ and Mike Stillman$^{c}$}}

\end{center}

\begin{center}

\textsl{$^{a}$Department of Physics, Cornell University, Ithaca, NY 14853, USA}\\
\textsl{$^{b}$Department of Physics, Northeastern University, Boston, MA 02115, USA}\\
\textsl{$^{c}$Department of Mathematics, Cornell University, Ithaca, NY 14853, USA}\\

\vspace{0.25cm}

%\vskip .3cm
\email{\tt md775@cornell.edu, co.long@northeastern.edu,\\ mcallister@cornell.edu, mike@math.cornell.edu }
\end{center}

\vspace{0.6cm}
\noindent

We show that the Weak Gravity Conjecture (WGC) implies a nontrivial upper bound on the volumes of the minimal-volume cycles in certain homology classes that admit no calibrated representatives.  In compactification of type IIB string theory on an orientifold $X$ of a Calabi-Yau threefold, we consider a homology class $[\Sigma] \in H_4(X,\mathbb{Z})$ represented by a union $\Sigma_{\cup}$ of holomorphic and antiholomorphic cycles.  The instanton form of the WGC applied to the axion charge $[\Sigma]$ implies an upper bound on the action of a non-BPS Euclidean D3-brane wrapping the minimal-volume representative $\Sigma_{\mathrm{min}}$ of $[\Sigma]$.  We give an explicit example of an orientifold $X$ of a hypersurface in a toric variety, and a hyperplane $\mathcal{H} \subset H_4(X,\mathbb{Z})$, such that for any $[\Sigma] \in H$ that satisfies the WGC, the minimal volume obeys $\mathrm{Vol}(\Sigma_{\mathrm{min}}) \ll \mathrm{Vol}(\Sigma_{\cup})$: the holomorphic and antiholomorphic components recombine to form a much smaller cycle.
In particular, the sub-Lattice WGC applied to $X$ implies large recombination, no matter how sparse the sublattice.
Non-BPS instantons wrapping $\Sigma_{\mathrm{min}}$ are then more important than would be predicted from a study of BPS instantons wrapping the separate components of $\Sigma_{\cup}$.
Our analysis hinges on a novel computation of effective divisors in $X$ that are not inherited from effective divisors of the toric variety.

\noindent
\vspace{2.1cm}

\noindent\today

\end{titlepage}
\tableofcontents\newpage

\section{Introduction}

In quantum theories of extended objects, such as string theories, there can be contributions to the path integral from extended objects wrapping cycles in spacetime.
Famous examples include worldsheet instantons, in which the Euclidean worldsheet of a string wraps a two-cycle, and D$p$-brane instantons, in which the Euclidean worldvolume of a D$p$-brane wraps a $p+1$-cycle.

Computing instanton contributions to the four-dimensional action is an important problem, doubly so in
theories in which the leading interactions of a given type are produced by instantons.
For example,
in a compactification of type IIB string theory on an orientifold $X$ of a Calabi-Yau threefold, the axions that result from reduction of the Ramond-Ramond four-form $C_4$ on four-cycles in $X$ have no non-derivative interactions at any order in perturbation theory, but acquire a potential from Euclidean D3-branes wrapping four-cycles \cite{Witten:1996bn}.

The semiclassical action of a Euclidean D3-brane includes a term proportional to the volume of the wrapped cycle, and so the Euclidean D3-brane will wrap a cycle that is at least locally volume-minimizing in its homology class.\footnote{We neglect for the moment the effects of the worldvolume gauge field, cf.~\cite{MMMS}.}
Thus, to compute instanton effects in the four-dimensional action, a first step is to compute the volumes of minimal cycles.  That is, one should find, for each class $[\Sigma] \in H_4(X,\mathbb{Z})$,
the
minimum-volume representative $\Sigma_{\mathrm{min}} \in [\Sigma]$, as well as its volume $\mathrm{Vol}(\Sigma_{\mathrm{min}})$.
This is still a far cry from an exact computation of the potential from instantons, which would require a
more refined study of the other degrees of freedom on the instanton worldvolume, including a counting of fermion zero modes.  Even so, computing $\mathrm{Vol}(\Sigma_{\mathrm{min}})$ gives insight about the relative importance of \emph{possible} instanton contributions.

When $X$ is a
K\"ahler threefold,
the computation of $\mathrm{Vol}(\Sigma_{\mathrm{min}})$ is almost trivial for a special set of classes $[\Sigma] \in H_4(X,\mathbb{Z})$: if $[\Sigma]$ can be represented by an effective divisor $\Sigma_\mathcal{E}$, then $\Sigma_\mathcal{E}$ is calibrated by the K\"ahler form $J$, and is absolutely volume-minimizing in its class, with volume
\begin{equation}\label{voleff}
\mathrm{Vol}(\Sigma_{\mathcal{E}}) = \mathrm{Vol}(\Sigma_{\mathrm{min}}) = \frac{1}{2}\int_{\Sigma_\mathcal{E}} J\wedge J\,,
\end{equation} which
is readily evaluated in terms of the K\"ahler parameters and intersection numbers of $X$.
One can similarly compute $\mathrm{Vol}(\Sigma_{\mathrm{min}})$ when $[\Sigma]$ can be represented by an anti-effective divisor\footnote{By an anti-effective divisor, we mean a finite formal sum of irreducible antiholomorphic hypersurfaces, with nonnegative integer coefficients.} $\Sigma_{\overline{\mathcal{E}}}$:
\begin{equation}\label{antieff}
\mathrm{Vol}(\Sigma_{\overline{\mathcal{E}}}) = \mathrm{Vol}(\Sigma_{\mathrm{min}}) = - \frac{1}{2}\int_{\Sigma_{\overline{\mathcal{E}}}} J\wedge J\,.
\end{equation}
Thus, for any effective or anti-effective divisor, the semiclassical action is easily computed, in the above approximation.

However, many classes in $H_4(X,\mathbb{Z})$ are neither effective nor anti-effective!
Computing
the minimum-volume representative is then highly nontrivial: in fact, it is an instance of one of the fundamental problems in geometric measure theory, the \emph{Plateau problem}, which we now review in slightly more general terms.

Suppose that one is given a Riemannian manifold $M$ of real dimension $n$, with fixed metric $g$. Given a homology class $[\Sigma] \in H_{p}(M,\mathbb{Z})$, $0<p<n$, what representative of $[\Sigma]$ has minimal volume?  A crucial subtlety is that the volume functional may not attain a minimum on any \emph{smooth} representative of $[\Sigma]$.  To make the variational problem well-posed, one should --- heuristically --- include appropriate limit points corresponding to mildly singular representatives.

The modern theory of this problem was founded by Federer and Fleming \cite{FF}.  They defined integral $p$-currents, which roughly correspond to formal sums of $p$-dimensional submanifolds, except for sets of $p$-dimensional Hausdorff measure zero.  Federer and Fleming showed that for any $M$ and $[\Sigma]$, there exists an integral $p$-current $\Sigma_{\mathrm{min}}$ representing $[\Sigma]$ that has minimal volume.  With some laxity of language, we refer to this minimal current as a minimal cycle.
Applied to string theory compactified on $M$, the minimal cycle $\Sigma_{\mathrm{min}}$ plausibly describes the configuration of a Euclidean brane in the class $[\Sigma]$, up to corrections from worldvolume fields beyond the embedding coordinates.

Geometric measure theory thus provides a sound framework for analyzing volume-minimization.
Even so, actually computing the volume $\mathrm{Vol}(\Sigma_{\mathrm{min}})$ of the minimal cycle in a given class remains difficult for general $M$ and $[\Sigma]$.  One knows that $\Sigma_{\mathrm{min}}$ exists \cite{FF}, and much is known about its degree of singularity \cite{Almgren}, but not much more can be said at this level of generality.
However, if $M$ is a K\"ahler
manifold,
the minimal cycles in effective classes are readily obtained from calibration data, as reviewed above.
One might therefore hope to characterize minimal cycles in general classes in terms of the properties of calibrated minimal cycles in effective classes.

Suppose, then, that $X$ is a K\"ahler threefold with $h^{2,0}(X)=0$, such as a Calabi-Yau threefold or an orientifold thereof.\footnote{In more general K\"ahler manifolds $X$ with $h^{2,0}(X)  \neq 0$, our analysis is only relevant to cycles dual to elements in the N\'{e}ron-Severi lattice of $X$.}  Then any $[\Sigma] \in H_4(X,\mathbb{Z})$ can be represented by a union of irreducible holomorphic and antiholomorphic four-cycles.  We refer to such a representative $\Sigma_{\cup}$ as a \emph{piecewise-calibrated} representative, because each irreducible component is either calibrated by $J$, or would be calibrated by $J$ following a reversal of orientation, cf.~\eqref{antieff}.
The volume $\mathrm{Vol}(\Sigma_{\cup})$ is then the sum of the volumes of the constituent holomorphic and antiholomorphic four-cycles, with each of these obtained by applying \eqref{voleff} or \eqref{antieff}, respectively.\footnote{In general, a class $[\Sigma]$ may have multiple piecewise-calibrated representatives $\{\Sigma_{\cup}^{I}\}$, with $I$ an index set, as we will explain further below.  By $\Sigma_{\cup}$ we mean the one of these that has the smallest volume.}
Thus, for given $[\Sigma]$, $\mathrm{Vol}(\Sigma_{\cup})$ is readily computed in terms of
the K\"ahler parameters of $X$.

To connect to the problem of Euclidean D3-branes, we note that in compactifications preserving $\mathcal{N}=1$ supersymmetry, only Euclidean D3-branes wrapping effective divisors can contribute to the superpotential \cite{Witten:1996bn}, and so the cycle volumes associated to superpotential terms are given in terms of calibration data.  On the other hand, a non-BPS Euclidean D3-brane wrapping the minimal-volume representative $\Sigma_{\mathrm{min}}$ of a non-effective class $[\Sigma]$ could contribute to the K\"ahler potential or to a higher F-term \cite{Beasley:2005iu}, depending on the number of fermion zero modes~\cite{Cvetic:2008ws, GarciaEtxebarria:2008pi, Buican:2008qe, Blumenhagen:2009qh}.
The piecewise-calibrated representative $\Sigma_{\cup}$ can be understood as an unstable collection of BPS and anti-BPS Euclidean D3-branes that can
recombine and fuse, reducing their volume, until they arrive at the non-BPS volume-minimizing configuration $\Sigma_{\mathrm{min}}$.

If we define the \emph{recombination fraction}
\begin{equation}\label{recf}
  \mathfrak{r}_{\Sigma} := \frac{\mathrm{Vol}(\Sigma_{\cup})-\mathrm{Vol}(\Sigma_{\mathrm{min}})}{\mathrm{Vol}(\Sigma_{\mathrm{min}})}\,,
\end{equation}
then $\mathrm{Vol}(\Sigma_{\cup})$ gives a useful approximation to $\mathrm{Vol}(\Sigma_{\mathrm{min}})$ if and only if $\mathfrak{r}_{\Sigma} \ll 1$.
When $\mathfrak{r}_{\Sigma} \gtrsim 1$, the minimal cycle and piecewise-calibrated cycle are quite different, and the action of a non-BPS instanton cannot be accurately estimated by adding up the action of its piecewise-calibrated BPS and anti-BPS `constituents'.
When $\mathfrak{r}_{\Sigma} \gg 1$, we say that \emph{large recombination} has occurred.

For the purpose of controlling the $\alpha'$ expansion in string theory, it is quite important to know whether $\mathrm{Vol}(\Sigma_{\mathrm{min}}) \approx
\mathrm{Vol}(\Sigma_{\cup})$ is a fair approximation, both for $[\Sigma] \in H_4(X,\mathbb{Z})$ as we have discussed, and for the analogous situation when $[\Sigma] \in H_2(X,\mathbb{Z})$ is a non-holomorphic curve class.  The reason is as follows.
For $X$ an orientifold of a Calabi-Yau threefold, let $\{\sigma_i\},~i=1,\ldots,h^{1,1}(X)$, be a basis of $H_2(X,\mathbb{Z})$, and define the K\"ahler parameters $t_i := \int_{\sigma_i} J$.
Many authors take $t_i \gg 1$ to be a \emph{sufficient} condition for ensuring that $X$ is large enough so that perturbative and nonperturbative corrections in the $\alpha'$ expansion are well-controlled.  There is some motivation for this condition: the actions of worldsheet instantons and Euclidean D3-branes wrapping effective curve and divisor classes, respectively, are indeed determined by the $t_i$ and by the topological data of $X$.  Likewise, for non-effective classes $[\Sigma]$, the volume $\mathrm{Vol}(\Sigma_{\cup})$ of the piecewise-calibrated representative is again determined by the $t_i$.  However, when $\mathfrak{r}_{\Sigma} \gg 1$ for some $[\Sigma]$, the corresponding non-BPS instanton is far more important than the computation of $\mathrm{Vol}(\Sigma_{\cup})$ would suggest.

To put this in stark terms, one can envision a `vulnerable' Calabi-Yau threefold $X$ for which, even with all $t_i \gg 1$, there are non-BPS instantons
with actions of order unity, even though all BPS instantons have large action.
The resulting corrections would invalidate the $\alpha'$ expansion.

Whether such vulnerable threefolds actually exist
is a purely geometric question, amounting to the existence of a pair $\{X,[\Sigma]\}$ with $\mathfrak{r}_{\Sigma}$ sufficiently large.
In principle this could be answered within geometric measure theory, without any input from physics.
However, to the best of our knowledge, this question is not settled for Calabi-Yau $n$-folds
for any $n$.  Perhaps the closest result is this: Micallef and Wolfson have exhibited a pair $\{X,[\Sigma]\}$ with $X$ a K3 surface near an orbifold limit, and $[\Sigma] \in H_2(X,\mathbb{Z})$, for which $\mathfrak{r}_{\Sigma} \ge \varepsilon$, with $\varepsilon$ parameterizing the distance from the orbifold limit \cite{Micallef}.
However, this establishes only that $\mathfrak{r}_{\Sigma} > 0$ can occur in a K3 surface, not that $\mathfrak{r}_{\Sigma} \gg 1$ can occur there.  In other words, Micallef and Wolfson proved that nonzero recombination can occur, not that large recombination can occur.\footnote{See also, for example,~\cite{federer1969geometric,Lawson,Chang,MR1216574,AREZZO2005209,ArezzoSun,Foscolo} for related ideas and further references.}

The goal of this work is to derive results about minimal surfaces that follow from the \emph{Weak Gravity Conjecture} (WGC) \cite{ArkaniHamed:2006dz}.
The instanton form of the WGC applied to an axion charge vector $\vec{Q}$ asserts an upper bound on the action $S_{\mathrm{min}}[\vec{Q}]$ of the smallest-action instanton of charge $\vec{Q}$, in terms of a certain quadratic norm $\|\vec{Q}\|$.
Schematically,
\begin{equation}
\mathrm{WGC} \Rightarrow S_{\mathrm{min}}[\vec{Q}] \le const. \times \|\vec{Q}\|\,.
\end{equation}
We present a precise version of this relation, applied to the case of interest, in \eqref{eq:fracvol2}.\footnote{The possibility of using the WGC to bound the volumes of non-holomorphic curves in compactifications on K3 surfaces was noted in \cite{Ooguri:2016pdq}.}

We will consider compactification of type IIB string theory on an orientifold $X$ of a complex threefold, and study the axions resulting from reduction of the Ramond-Ramond four-form $C_4$ on four-cycles.
For a given class $[\Sigma]$, the axion charge vector is then $[\Sigma]$ itself, understood as a vector in $H_4(X,\mathbb{Z})$.  The corresponding instantons are Euclidean D3-branes wrapping the minimal surface $\Sigma_{\mathrm{min}}$, with Euclidean action whose real part is proportional to $\mathrm{Vol}(\Sigma_{\mathrm{min}})$.  The quadratic norm $\|\Sigma\|$ is determined by the metric on K\"ahler moduli space, see \eqref{eq:wgc}.

In this situation, the WGC applied to the class $[\Sigma]$ asserts a lower bound
\begin{equation}\label{axwgsa}
\mathrm{WGC} \Rightarrow \mathfrak{r}_{\Sigma} \ge \mathfrak{r}_{\Sigma}^{\mathrm{min}}\,,
\end{equation} with $\mathfrak{r}_{\Sigma}^{\mathrm{min}}$ computable in terms of $\|\Sigma\|$, cf.~\eqref{eq:fwgc}.
We will give an explicit example of an orientifold $X$ of a Calabi-Yau threefold hypersurface in a toric variety, and a hyperplane $\mathcal{H} \subset H_4(X,\mathbb{Z})$,
for which $\mathfrak{r}_{\Sigma}^{\mathrm{min}} \gg 1$ for all $[\Sigma] \in H$.
Thus, if the WGC applies to \textit{any} $[\Sigma] \in H$, it implies $\mathfrak{r}_{\Sigma} \gg 1$.  In particular, if the sub-Lattice WGC \cite{Heidenreich:2016aqi}
holds in $X$, it follows that $\mathfrak{r}_{\Sigma} \gg 1$, and so certain four-cycles necessarily undergo large recombination.

The evidence that the sub-Lattice WGC is a fact about quantum gravity is not conclusive, and for the purpose of this work we are agnostic about its truth value.
Our result can on the one hand be read as preparing a purely geometric test, the condition $\mathfrak{r}_{\Sigma} \ge \mathfrak{r}_{\Sigma}^{\mathrm{min}}$ of \eqref{axwgsa}, failure of which in a single example of a string compactification would disprove the sub-Lattice WGC.  It would be striking if the tools of geometric measure theory could be used in this way to discern properties of quantum gravity.  On the other hand, if the sub-Lattice WGC is taken to be true, either provisionally or based on further evidence about quantum gravity, then our result can be read as an upper bound on the volumes of certain minimal surfaces.

The organization of this paper is as follows.  In \S\ref{axionwgc} we collect the necessary results about the WGC.
In \S\ref{sec::examples} we present an explicit example where $\mathfrak{r}_{\Sigma}^{\mathrm{min}} \gg 1$. We discuss some implications, and conclude, in \S\ref{disc}.
Appendix \ref{sec:app} gives more details about the geometry of the example. In Appendix \ref{sec:effective}, we prove a useful lemma about effective divisors in the example.

\section{Axions, Orientifolds, and Weak Gravity}\label{axionwgc}

We begin by explaining how the  WGC constrains the volumes of non-holomorphic four cycles. Let us first describe our normalization conventions.  We define the string length $\ell_s$ as
\begin{equation}
	\ell_s = 2 \pi \sqrt{\alpha'}.
\end{equation}
Following the conventions of \cite{DDFGK} we take
\begin{equation}
	\frac{1}{2 \kappa_{10}^2} = \frac{2 \pi}{\ell_s^8}, \quad T_p = \frac{2 \pi}{g_s \ell_s^{p+1}},
\end{equation}
where $T_p$ is the tension of a Dp-brane. We measure all cycle volumes using the ten-dimensional Einstein-frame metric, and express these volumes in units of $\ell_s$. The four-dimensional Planck mass is given by
\begin{equation}
	M_{\text{pl}}^2 = \frac{4\pi \cV}{\ell_s^2 g_s^{1/2}},
\end{equation}
where $\cV$ is the compactification volume. Henceforth, we set $\ell_s=1$, so that the only dimensionful parameter appearing in our equations is $M_{\text{pl}}$.

We consider type IIB string theory compactified on an O3/O7 orientifold $X$ of a Calabi-Yau threefold $\widetilde{X}$.\footnote{In this case we have $h^{2,0}(X)  = 0$, see e.g.~\cite{Weigand:2018rez}, and therefore all elements of $H_4(X,\mathbb{Z})$ admit piecewise-calibrated representatives.}
Given a basis $\{D_i\}$, $i=1 \dots h^{1,1}$ for $H_4(X,\mathbb{Z})$, the K\"ahler moduli are written as
\begin{equation}
T^i = \frac{1}{2}\int\limits_{D_i} J \wedge J + i\, \int\limits_{D_i} C_4 \equiv \tau^i + i\theta^i\, ,  \quad i = 1\ldots h^{1,1}\, ,
\end{equation}
where $C_4$ is the Ramond-Ramond four-form. The effective Lagrangian for the axions $\theta^i$ has the form
\begin{equation}\label{eq:potential}
\cL =\frac{M_\text{pl}^2}{2} \mathcal{R}_4 -\frac{M_\text{pl}^2}{2} K_{ij}\partial^{\mu}\theta^i\partial_{\mu}\theta^j - V(\theta).
\end{equation}
At large volume, the K\"ahler metric $K_{ij}$ on the complexified K\"ahler moduli space of $X$ is independent of the $\theta^i$ and is obtained from the tree level K\"ahler potential $\cK= -2 \log{\cV}$.
In perturbation theory, the axions $\theta^i$ enjoy continuous shift symmetries due to the ten-dimensional gauge invariance of $C_4$. These are broken to discrete shift symmetries by non-perturbative contributions to the scalar potential. The discrete shift symmetries generate the period lattice, denoted $\Gamma^*$, so the general axion potential can be written as
\begin{equation}
V(\theta) = \sum\limits_{[\Sigma] \in \Gamma^{*}} Z_{\Sigma}e^{2 \pi i [\Sigma]\cdot\vec{\theta}}\, ,
\end{equation}
where  $Z_{-\Sigma} = Z_{\Sigma}^{*}$ as $V(\theta)$ is real.  We will consider the potential $V(\theta)$ generated by Euclidean D3-branes, which wrap minimum-volume representatives $\Sigma_{\mathrm{min}}$ of homology classes $[\Sigma] \in H_4(X,\mathbb{Z})$,
and thus $\Gamma^{*} = H_4(X,\mathbb{Z})$.
In the semiclassical regime, the coefficients $Z_{\Sigma}$ are given by
\begin{equation}
Z_{\Sigma} = A_{\Sigma} e^{-S_{\Sigma}}\, ,
\end{equation}
where $A_{\Sigma}$ is the one-loop determinant and the semiclassical action $S_{\Sigma}$ of the instanton is determined by the cycle volume,
\begin{equation}\label{sisvol}
\mathrm{Re}\,S_{\Sigma} = 2\pi \mathrm{Vol}(\Sigma_{\mathrm{min}})\, ,
\end{equation}
measured with respect to the metric $g$ on $X$.

The instanton form of the WGC applied to a charge vector $[\Sigma]$ states that there exists an instanton with charge $[\Sigma]$ that satisfies
\begin{equation}\label{eq:wgc}
\mathrm{Re}\,S_\Sigma \leq \awgc\,\|\Sigma\| , \qquad \text{where} \quad \|\Sigma\| := 2 \pi \sqrt{\Sigma_i (K^{-1})^{ij} \Sigma_j}\, ,
\end{equation}
for some $c \sim 1$.  Various forms of the WGC constrain various subsets of the charge lattice $H_4(X,\mathbb{Z})$.
In particular, the sub-Lattice WGC states that there exists a sublattice $\Gamma_\text{sub}^{*} \subseteq \Gamma^{*}$ of finite index\footnote{The index of a sublattice $\Gamma_\text{sub}^{*}$ is the smallest integer $n$ such that $n[\Sigma] \in \Gamma_\text{sub}^{*} $ for all $ [\Sigma] \in \Gamma^{*}$.} such that at every site in $\Gamma_\text{sub}^{*}$ there is an instanton that satisfies \eqref{eq:wgc}.\footnote{The foundational paper on the Weak Gravity Conjecture is \cite{ArkaniHamed:2006dz}.  The sub-Lattice WGC was formulated by Heidenreich, Reece, and Rudelius in \cite{Heidenreich:2016aqi}, building on \cite{Heidenreich:2015nta},	and closely related work by Montero, Shiu, and Soler appears in \cite{Montero:2016tif}.  Other related recent developments include \cite{Cheung:2014vva,Rudelius:2015xta,Brown:2015iha, Brown:2015lia, Cottrell:2016bty, Heidenreich:2016aqi, Heidenreich:2017sim, Hamada:2017yji, Cheung:2018cwt, Andriolo:2018lvp, Heidenreich:2018kpg, Hamada:2018dde,Grimm:2018ohb,Lee:2018urn, Lee:2018spm}.}

Using \eqref{sisvol}, the WGC constraint \eqref{eq:wgc} becomes
\begin{equation}\label{eq:fracvol2}
\mathrm{Vol}(\Sigma_{\mathrm{min}})  \le \awgc\,\sqrt{\Sigma_i (K^{-1})^{ij} \Sigma_j}\,.
\end{equation}
Equivalently, using \eqref{recf}, the WGC implies a lower bound on the recombination fraction
\begin{equation}\label{eq:fwgc}
\mathfrak{r}_{\Sigma} \ge \mathfrak{r}^{\mathrm{min}}_{\Sigma} := \frac{2\pi}{\awgc}\frac{\mathrm{Vol}(\Sigma_{\cup})}{\|\Sigma\|} - 1\,.
\end{equation}

In the next section, we will present an example
in which --- as a purely geometric statement, making no assumption concerning the WGC --- for any charge vector $[\Sigma]$ lying in a specific hyperplane $\mathcal{H} \subset H_4(X,\mathbb{Z})$, the K\"ahler form $J$ can be chosen so as to make the lower bound $\mathfrak{r}^{\mathrm{min}}_{\Sigma}$ arbitrarily large.  If the WGC applies to any such $[\Sigma]$, it implies $\mathfrak{r}_{\Sigma} \gg 1$.

One might worry that $\mathfrak{r}_{\Sigma} \gg 1$ signals the breakdown of the $\alpha'$ expansion as the volumes of certain cycles become small. This is not the case: $\mathfrak{r}^\text{min}_{\Sigma}$ is unaffected by the scaling $J \rightarrow \lambda J$, for $\lambda \in \mathbb{R}^{+}$, and so it is always possible to rigidly dilate $X$ as much as desired without affecting the relation \eqref{eq:fwgc}.

Before proceeding to the example, let us first establish that for Euclidean D3-branes wrapping effective or anti-effective cycles, \eqref{eq:wgc} is always satisfied with $c = \sqrt{3}/2$. The proof is as follows. We expand the K\"ahler form $J$ in terms of the Poincar\'e duals $\omega_i$ of a set of basis divisors $D_i$,
\begin{equation}
J = t_i \omega_i\, .
\end{equation}
In terms of the K\"ahler parameters $t_i$, the divisor volumes and the volume of $X$ are
\begin{equation}\label{eq:vols}
\begin{aligned}
\tau_i &= \frac{1}{2} \kappa_{ijk} t_j t_k\, ,\\
\cV &= \frac{1}{6} \kappa_{ijk} t_i t_j t_k\, ,\\
\end{aligned}
\end{equation}
where $\kappa_{ijk} = \# D_i \cap D_j \cap D_k$ are the triple intersection numbers. The inverse K\"ahler metric $(K^{-1})^{ij}$ has the form
\begin{equation}\label{eq:kmet}
(K^{-1})^{ij} = 4 \tau_i \tau_j - 4 \cV \kappa_{ijk}t_k\,.
\end{equation}
Using \eqref{eq:vols} and \eqref{eq:kmet} we can write
\begin{equation}\label{therat}
 \frac{\mathrm{Re}\,S_\Sigma}{\|{\Sigma}\|} = \frac{2 \pi \text{Vol} (\Sigma)}{\|{\Sigma}\|} = \frac{2\pi t\cdot K^{-1}\cdot \Sigma}{8 \pi \cV \sqrt{\Sigma\cdot K^{-1} \cdot \Sigma}}  = \frac{t\cdot K^{-1}\cdot \hat{\Sigma}}{4\cV}\,,
\end{equation}
where we defined $\hat{\Sigma}=\Sigma/\| {\Sigma} \|$. The ratio in \eqref{therat} is maximized when $\Sigma \propto t$,
and since $t \cdot K^{-1} \cdot t = 12 \cV^2$ we have
\begin{equation}
 \frac{\mathrm{Re}\,S_\Sigma}{\|{\Sigma}\|} \leq \frac{12 \cV^2}{4\cV \sqrt{12 \cV^2}} = \frac{\sqrt{3}}{2} \, .
\end{equation}

\section{An Orientifold where Weak Gravity Implies Large Recombination}\label{sec::examples}
We now give an explicit example of an orientifold $X$ of a Calabi-Yau threefold hypersurface, and a hyperplane $\mathcal{H} \subset H_4(X,\mathbb{Z})$, such that for any $[\Sigma] \in H$, the ratio
$\mathfrak{r}_\Sigma^\text{min}$ defined in \eqref{eq:fwgc} can be made arbitrarily large by a choice of the K\"ahler form $J$.\footnote{Further details are presented in Appendix~\ref{sec:app}.}

We consider the product $V = \mathbb{P}^1 \times \mathbb{P}^1 \times \mathbb{F}_4$, where $\mathbb{F}_4$ is the fourth Hirzebruch surface.
Realized as a toric fourfold, $V$ has degrees in the Cox ring
\begin{equation}\label{eqn:charges2}
\begin{array}{|c|c|c|c|c|c|c|c|c|}
\hline
x_0 & x_1 & x_2 & x_3 & x_4 & h & x_6 & \eta\\
\hline
1 & 1 &0 & 0 & 0 &0 & 0 & 0\\
0 & 0 &1 & 1 & 0 &0 & 0 & 0\\
0 & 0 &0 & 0 & 0 &1 & 0 & 1\\
0 & 0 &0 & 0 & 1 &-4 & 1 & 0\\
\hline
\end{array}\
\end{equation}
and Stanley-Reisner ideal
\begin{equation}\label{eq:sri2}
\left< x_0 x_1, x_4 x_6, x_2 x_3, h \eta \right>\, .
\end{equation}
The orientifold $X \subset V$ is defined by the vanishing of a polynomial $F$ of the form
\begin{equation}\label{eqn:noncy}
F = h P_{(2,2,0,4)} - \eta  P_{(2,2,0,0)}\, ,
\end{equation}
where $P_{(2,2,0,4)}$ and $P_{(2,2,0,0)}$ are polynomials independent of $h$ and $\eta$.
We define a basis $\{D_a, D_b, D_c, D_d\}$ for $H_{4}(X,\mathbb{Z})$ with degrees
\begin{align}\label{eq:basis}
& [D_a] = (1,0,0,0)\, ,\nonumber \\
& [D_b] = (0,1,0,0)\, ,\nonumber \\
& [D_c] = (0,0,1,-4)\, ,\nonumber \\
& [D_d] = (0,0,0,1)\, ,
\end{align}
and expand the K\"ahler form $J$ as
\begin{equation}
J = t_a \omega_a  + t_b \omega_b  + t_c \omega_c + t_d \omega_d \, ,
\end{equation}
where $\omega_A$, $A \in \{a,b,c,d\}$, are Poincar\'e dual to $D_A$. The Mori cone of $V$ is simplicial, and the volumes of its generators are
\begin{equation}
\{t_a, t_b,t_c,t_d - 4t_c\}\, .
\end{equation}
We now make the notational replacement
\begin{equation}\label{defdelta}
t_d \rightarrow t_e + 4 t_c\,,
\end{equation}
so that the K\"ahler cone conditions read $t_a >0, t_b>0, t_c>0$, and $t_e > 0$. With this parameterization the volume $\cV$ of $X$ takes the form
\begin{equation}\label{eq:xvol}
\cV = 4 t_b t_c (t_a + t_c) + t_b(t_a + 2t_c)t_e + 2t_at_c(2t_c + t_e).
\end{equation}

To show that the WGC implies large recombination, we need to find a non-effective divisor class $[\Sigma] \in H_4(X, \mathbb{Z})$, and a point in the K\"ahler cone of $X$ for which $\mathfrak{r}_\Sigma^\text{min}$ is large. To find such a non-effective divisor class and characterize its piecewise calibrated representatives, we must first identify the effective and anti-effective divisor classes on $X$.

Recall that for $Y$ an algebraic $n$-fold, a homology class $[\Sigma] \in H_{2m}(Y,\mathbb{Z})$, for $1 \le m \le n$, is called (anti-)effective if it can be represented by a (anti-)holomorphic $2m$-cycle.  The cone of effective $2m$-cycles, which we denote by $\mathcal{E}_m(Y)$, is
by definition the cone in $H_{2m}(Y,\mathbb{R})$ generated by the effective classes in $H_{2m}(Y,\mathbb{Z})$.  Likewise, the cone $\overline{\mathcal{E}}_m(Y)$ of anti-effective $2m$-cycles is generated by the anti-effective classes in $H_{2m}(Y,\mathbb{Z})$.

When $Y$ is a hypersurface of dimension $n$ in a toric variety $V$ of dimension $n+1$, some effective divisors $D$ on $Y$ are inherited from effective divisors $\widehat{D}$ on $V$, with $D = Y \cap \widehat{D}$. Enumerating these inherited divisors is straightforward, because $\mathcal{E}_{n}(V)$ is readily computed from toric data.  However, a general such $Y$ may also contain \emph{autochthonous} effective divisors, i.e.~effective divisors $D$ that are not of the form $Y \cap \widehat{D}$ for any effective divisor $\widehat{D}$ on $V$, and so are not simply inherited from the divisors of $V$.  Certain autochthonous divisors can be identified by attempting to factor the defining equation of $Y$ \cite{autochpaper}.

Finding all autochthonous divisors on $X$ and computing the generators of $\mathcal{E}_2(X)$ exactly is difficult and beyond the scope of this work. Instead, we will construct a cone $\cE^{\text{out}}_2(X)$ that contains $\cE_2(X)$. We first note that any effective divisor in $X$ is an effective surface in $V$, i.e. $\mathcal{E}_2(X) \subseteq \mathcal{E}_2(V)$.\footnote{The converse is not true in general. We will show below that $\mathcal{E}_2(X) \neq \mathcal{E}_2(V)$ in our example.} Moreover, by a theorem of Fulton, MacPherson, Sottile, and Sturmfels~\cite{fulton}, the cone of effective $n$-cycles $\mathcal{E}_n(V)$ in $V$ is generated by the toric $n$-cycles of $V$. We consider a general line bundle $\mathcal{L}$ on $V$, and demand $\mathcal{L}|_X \in \mathcal{E}_2(V)$. We find\footnote{Further details will appear in \cite{autochpaper}.} that any such line bundle is generated over $\mathbb{Z}_{\ge0}$ by:
\begin{align}\label{eqn:outergens}
& [D_a] = (1,0,0,0)\, ,\nonumber \\
& [D_b] = (0,1,0,0)\, ,\nonumber \\
& [D_c] = (0,0,1,-4)\, ,\nonumber \\
& [D_d] = (0,0,0,1)\, ,\nonumber \\
& [D_e] = (-2,2,1,0)\, ,\nonumber \\
& [D_f] = (2,2,-1,4)\, ,\nonumber \\
& [D_g] = (2,-2,1,0)\, .
\end{align}
The divisors $[D_a],\ldots,[D_g]$ are the extremal generators of $\cE^{\text{out}}_2(X)$.
Their volumes are
\begin{align}\label{eq:divvols}
 \tau_a &= 4t_bt_c + 4t_c^2 + t_bt_e + 2t_ct_e\, ,\nonumber \\
 \tau_b &= 4t_at_c + 4t_c^2 + t_at_e + 2t_ct_e\, ,\nonumber \\
 \tau_c &= 2t_at_e + 2t_bt_e\, ,\nonumber \\
 \tau_d &= t_at_b + 2t_bt_c + 2t_at_c\, ,\nonumber \\
 \tau_e &= 4t_at_b + 16t_at_c + 4t_at_e\, ,\nonumber \\
 \tau_f &= 8t_at_c + 8t_bt_c + 16t_c^2 + 8t_ct_e\, ,\nonumber \\
 \tau_g &=  4t_at_b + 16t_bt_c + 4t_bt_e\, .
\end{align}

Next, we consider the hyperplane $\mathcal{H} \subset H_4(X,\mathbb{Z})$ spanned by charge vectors of the form
\begin{equation}
[\Sigma^{(k)}] := (k,-k,c,d)\,,
\end{equation}
where $k \in \mathbb{Z}_{+}$ and $c,d \in \mathbb{Z}$. In Appendix \ref{sec:effective}, we prove that every such $[\Sigma^{(k)}]$ is neither effective nor anti-effective. We will now demonstrate that any WGC that constrains a class $[\Sigma^{(k)}]$ implies large recombination.

As $[\Sigma^{(k)}]$ is non-effective, we express it as a sum of effective and anti-effective divisor classes, and construct a piecewise-calibrated representative $\Sigma_{\cup}^{(k)}$:
\begin{alignat}{2}
	[\Sigma^{(k)}] &= \sum_{[\Sigma_i] \in \cE_2(X)} \alpha_i [\Sigma_i] \, &&+ \, \sum_{[\overline{\Sigma}_i] \in \overline{\cE}_2(X)} \beta_i [\overline{\Sigma}_i]\, , \label{eq:decomp1} \\
	\Sigma_{\cup}^{(k)} &= \bigcup_{[\Sigma_i] \in \cE_2(X)} \alpha_i \Sigma_i \, &&\cup \, \bigcup_{[\overline{\Sigma}_i] \in \overline{\cE}_2(X)} \beta_i \overline{\Sigma}_i  \, , \label{eq:decomp2}
\end{alignat}
where $\alpha_i, \beta_i \in \mathbb{Z}_{\geq 0}$. The volume of $\Sigma_\cup^{(k)}$ is the sum of the volumes of its holomorphic and anti-holomorphic constituents,
\begin{equation}\label{eq:volsum}
	\text{Vol}(\Sigma_{\cup}^{(k)}) = \sum_{[\Sigma_i] \in \cE_2(X)} \alpha_i \text{Vol}(\Sigma_i) \, + \, \sum_{[\overline{\Sigma}_i] \in \overline{\cE}_2(X)} \beta_i \text{Vol}(\overline{\Sigma}_i) \, .
\end{equation}
Note that the expansion \eqref{eq:decomp1} is not unique: any given $[\Sigma^{(k)}]$ can be expanded in infinitely many inequivalent ways. We define $\Sigma_{\cup}^{(k)}$ to be the piecewise-calibrated representative with the smallest volume. In general, finding the smallest-volume piecewise-calibrated representative is difficult.
However, we will work in a region of the K\"ahler cone where the problem is simplified considerably. Consider the locus where $t_a = t_b = \sigma$, $t_c = t_e = \delta$ for some $\delta, \sigma >0$. The volumes of the generators of $\cE^{\text{out}}_2 (X)$, cf. \eqref{eq:divvols}, become
\begin{align}\label{eq:divvols2}
\tau_a &=6\delta^2 + 5\sigma \delta \, ,\nonumber \\
\tau_b &=6\delta^2 + 5\sigma \delta \, ,\nonumber \\
\tau_c &= 4\sigma \delta\, ,\nonumber \\
\tau_d &= 4\sigma \delta + \sigma^2 \, ,\nonumber \\
\tau_e &= 20\sigma \delta + 4\sigma^2 \, ,\nonumber \\
\tau_f &= 24\delta^2 + 16\sigma \delta \, ,\nonumber \\
\tau_g &= 20\sigma \delta + 4\sigma^2 \, .
\end{align}
The quadratic norm $\|\Sigma^{(k)}\|$, cf.~\eqref{eq:wgc}, is computed using \eqref{eq:vols}, \eqref{eq:kmet} and \eqref{eq:xvol}:\footnote{Intersection numbers $\kappa_{ijk}$ can be calculated by differentiating \eqref{eq:xvol}.}
\begin{align}
\|{\Sigma}^{(k)}\|^2 &= 8 \pi^2 \bigg(60 k^2 \delta^3 \sigma + (128 c^2 + 16 c d + 32d^2 + 25k^2)\delta^2 \sigma^2 \nonumber \\
&+ (40c^2 - 4c d + 16d^2)\delta \sigma^3 + 4d^2 \sigma^4\bigg)\,.
\end{align}

We will now show that the volume of any piecewise-calibrated representative of $[\Sigma^{(k)}]$ obeys $\text{Vol}( \Sigma_{\cup}^{(k)}) > 12 \delta^2$.
As $\cE_2(X) \subset \cE^{\text{out}}_2 (X)$, the volume of any effective or anti-effective divisor can be written as a nonnegative integer sum of the volumes given in \eqref{eq:divvols}.  We can therefore write
\begin{equation}
\text{Vol}(\Sigma_{\cup}^{(k)}) = \sum_A \gamma^A \tau_A\,,
\end{equation}
where $A \in \{a,b,c,d,e,f,g\}$ and $\gamma^A \in \mathbb{Z}_{\geq 0}$.
Comparing to \eqref{eq:divvols}, we can write
\begin{equation}
\text{Vol}(\Sigma_{\cup}^{(k)}) = m_1 \delta^2 + m_2 \delta \sigma + m_3 \sigma^2,
\end{equation} with $m_1, m_2, m_3 \in \mathbb{Z}_{\ge 0}$.
Let us now show that $m_1 \ge 12$. We can represent each homology class by the corresponding degrees, as in \eqref{eq:basis}, and rewrite the sum \eqref{eq:decomp1} as
\begin{equation}\label{genexp}
[\Sigma^{(k)}] = \sum_{i} \mathcal{C}^{(k)}_i\,(a_i,b_i,c_i,d_i)
\end{equation} where $i$ runs over effective and anti-effective divisors, and $\mathcal{C}^{(k)}_i \in \mathbb{Z}_{\geq 0}$.
We then break up the general sum \eqref{genexp} into three subclasses:
\begin{align}
[\Sigma^{(k)}] = &\sum_{a_i = b_i = 0} A^{(k)}_i (a_i,b_i,c_i,d_i)\, \nonumber \\
+ &\sum_{a_i = -b_i \neq 0} B^{(k)}_i (a_i,b_i,c_i,d_i)\, \nonumber \\
+ &\sum_{a_i \neq -b_i} C^{(k)}_i (a_i,b_i,c_i,d_i),
\end{align}
where $A_i^{(k)}, B_i^{(k)}, C_i^{(k)} \in \mathbb{Z}_{\geq 0}$.  By Theorem 1 of Appendix \ref{sec:effective}, any divisor class with $a_i = -b_i \neq 0$ is non-effective and cannot contribute to the sum, so we have $B_i^{(k)} = 0$. By assumption $k \neq 0$, so there must be at least one term where $a_i \neq 0$, and at least one other term where $b_i \neq 0$: that is, $C_i^{(k)} \neq 0$ for at least two values of $i$.  From \eqref{eq:basis} and \eqref{eq:divvols2}, we see that the volume of any divisor with $a_i \neq -b_i$ is greater than $6 \delta^2$.  It follows that $\text{Vol}(\Sigma_{\cup}^{(k)}) > 12 \delta^2$.

The quadratic norm $\|{\Sigma}^{(k)}\|$ grows only as $\cO(\delta^{3/2})$ in the limit $\delta \gg \sigma$.
Thus, for $\delta \gg \sigma$, there is a lower bound on the recombination fraction:
\begin{equation}
	\mathfrak{r}^{\text{min}}_{\Sigma^{(k)}} = \frac{2\pi}{\awgc}\frac{\mathrm{Vol}(\Sigma^{(k)}_{\cup})}{\|\Sigma^{(k)}\|} - 1 \sim \cO(\delta^{1/2})\,.
\end{equation}
By taking $\delta$ large while holding $\sigma$ fixed, we can make $\mathfrak{r}^{\text{min}}_{\Sigma^{(k)}}$ as large as desired.

As a concrete example, take $\delta=100$, $\sigma=1$ and $\Sigma=(-k,k,0,0)$, for some $k \in \mathbb{Z}_{\geq 0}$. We find $\| \Sigma \| = 1000 \pi k\sqrt{482}$ and $\text{Vol}(\Sigma_{\cup}^{(k)}) > 120000k$.  Setting $c=1$ in \eqref{eq:wgc}, we get
\begin{equation}
	\mathfrak{r}^{\text{min}}_{\Sigma^{(k)}} > 9.9\,.
\end{equation}

The argument presented above applies for any $k \in \mathbb{Z}_+$ and any $c,d \in \mathbb{Z}$. Thus, any version of the WGC that constrains at least one charge vector lying in the hyperplane $\mathcal{H}$ spanned by the $[\Sigma^{(k)}]$ necessarily implies that large recombination occurs.  In particular, because the vector $(1,-1,c,d)$ lies in $\Gamma^{*}$,
for any sublattice $\Gamma_{\text{sub}}^{*} \subset \Gamma^{*}$ there exists a finite $n \in \mathbb{Z}_{+}$ such that $(n,-n,nc,nd) \in \Gamma_{\text{sub}}^{*}$.  Thus, the sub-Lattice WGC for any sublattice $\Gamma_{\text{sub}}^{*}$, no matter how sparse, implies that some of the charges in the hyperplane spanned by $[\Sigma^{(k)}]$ must obey \eqref{eq:wgc}.  It follows that the sub-Lattice WGC requires large recombination in our example.

\section{Discussion}\label{disc}

We have argued that if the Weak Gravity Conjecture is true, the volumes of certain minimal surfaces must obey a nontrivial upper bound.
The bound requires that each minimal surface is sufficiently small compared to any union of holomorphic and antiholomorphic surfaces representing the same homology class.

Specifically, in a compactification of type IIB string theory on an O3/O7 orientifold of any Calabi-Yau threefold $X$, and for any $[\Sigma] \in H_4(X,\mathbb{Z})$ that is neither effective nor anti-effective on $X$, but is constrained by the WGC, the volume $\mathrm{Vol}(\Sigma_{\mathrm{min}})$ of the minimal-volume representative of $[\Sigma]$ must obey \eqref{eq:fracvol2}.  This amounts to the statement that the minimal cycle representing $[\Sigma]$ must be sufficiently small in comparison to any union of holomorphic and antiholomorphic cycles representing $[\Sigma]$.

The outline of our argument was as follows.  The real part of the action of a Euclidean D3-brane in a class $[\Sigma]$, ignoring the effects of magnetization and of the fluctuation determinant, is $2\pi\mathrm{Vol}(\Sigma_{\mathrm{min}})$.
The instanton form of the WGC then gives an upper bound on $\mathrm{Vol}(\Sigma_{\mathrm{min}})$ in terms of the quadratic norm $\|\Sigma\|$, cf.~\eqref{eq:wgc}, of the axion charge $[\Sigma]$.
Although $\mathrm{Vol}(\Sigma_{\mathrm{min}})$ is difficult to compute, one can instead compare $\|\Sigma\|$ to the volume $\mathrm{Vol}_\cup(\Sigma)$ of a piecewise-calibrated representative of $[\Sigma]$, which corresponds
to the sum of the actions of a collection of BPS and anti-BPS Euclidean D3-branes whose total charge is $[\Sigma]$.
Given any such piecewise-calibrated representative, one can think of fusing and recombining the holomorphic and antiholomorphic constituents to form a non-holomorphic cycle with smaller total volume. Any version of the WGC that nontrivially constrains $[\Sigma]$ asserts that
$\mathrm{Vol}(\Sigma_{\mathrm{min}})<\mathrm{Vol}_\cup(\Sigma)$, with \eqref{eq:fwgc} giving the precise relation.

The physical origin of our statement is simply that the WGC prescribes a minimum amount of binding energy that must be released when certain collections of BPS and anti-BPS states bind to form a non-BPS state.  We have mapped this requirement into a statement about minimal surfaces in orientifolds of Calabi-Yau threefolds, i.e.~about an infinite class of instances of the Plateau problem.

We should comment that if one is given the topological data of an orientifold $X$ of a Calabi-Yau threefold, together with the K\"ahler cone $\mathcal{K}(X)$ of $X$ and the cone $\mathcal{E}_2(X)$ of effective divisors on $X$, it is straightforward to evaluate $\mathrm{Vol}_\cup(\Sigma)$ and $\|\Sigma\|$ at any point in $\mathcal{K}(X)$.
Even so, it turned out to be nontrivial to exhibit such an $X$, as an orientifold of a hypersurface in a toric variety $V$, for which it was possible to show $\mathrm{Vol}_\cup(\Sigma) \gg \|\Sigma\|$.
The catch is that $\mathcal{E}_2(X)$ is \emph{not} simply inherited from $V$ by intersecting $X$ with divisors of $V$ --- see \cite{Demirtas:2018akl} for a related discussion --- and with incomplete knowledge of $\mathcal{E}_2(X)$ it is difficult to compute $\mathrm{Vol}_\cup(\Sigma)$. We overcame this limitation by showing, in Appendix \ref{sec:effective}, that a particular family of divisor classes are neither effective nor anti-effective, and so arrived at the example of \S\ref{sec::examples}.

Some cautionary remarks are necessary. We have ignored degrees of freedom on the Euclidean D3-brane worldvolume other than the embedding coordinates, we have omitted the Pfaffian prefactor, and we have ignored the effects of magnetization.  These simplifying assumptions led to \eqref{sisvol}, which we used to relate the defining statement \eqref{eq:wgc} of the instanton WGC to the version \eqref{eq:fracvol2} used in our analysis.  It is therefore logically possible that there exist examples of pairs $\{X,[\Sigma]\}$ in which $\mathrm{Vol}(\Sigma_{\mathrm{min}})$ violates \eqref{eq:fracvol2}, and yet the WGC \eqref{eq:wgc} holds nonetheless, because \eqref{sisvol} is strongly violated.  We find such large violations of \eqref{sisvol} to be implausible in the regime where all K\"ahler parameters are large, but this deserves more detailed study.

We have also assumed, as in most of the literature, that the WGC applies at generic points in the moduli space of $X$. It might be that the WGC only applies on-shell, i.e., at the discrete set of points in the moduli space where the scalar potential is minimized. One would then need to  stabilize the moduli, and check the relation \eqref{eq:fwgc} at the vacua of the resulting effective field theory.

This work opens the possibility of proving or disproving various versions of the WGC using the methods of geometric measure theory. If one were to find a violation of \eqref{eq:fracvol2} in a single pair $\{X,[\Sigma]\}$, this would be incompatible with any WGC that constrains $[\Sigma]$.

Another important application concerns the validity of the $\alpha'$ expansion.
As we explained in the introduction, a common assumption is that $\alpha'$ corrections to the effective action are small when the K\"ahler form $J$ is `big' enough so that every holomorphic curve and holomorphic divisor is large in string units.
However, if $\mathfrak{r}_{\Sigma}$ is sufficiently large for some $[\Sigma]$, then a non-BPS instanton with charge $[\Sigma]$ can give corrections to the effective action that are large enough to invalidate the $\alpha'$ expansion.  One should therefore ask whether there exist pairs $\{X,[\Sigma]\}$ with dangerously large $\mathfrak{r}_{\Sigma}$.

It would also be interesting to understand the degree to which the WGC implies recombination in other types of four-dimensional $\mathcal{N} = 1$ compactifications, such as type IIA string theory on a Calabi-Yau orientifold, or M-theory on a $G_2$ manifold.

More broadly, we anticipate that geometric measure theory can be used to illuminate the study of non-BPS instantons and of the Weak Gravity Conjecture.

\section*{Acknowledgments}

We thank Chris Beasley, Naomi Gendler, Jim Halverson, Ben Heidenreich, Manki Kim,  Tom Rudelius, John Stout, Ben Sung, Irene Valenzuela, and Max Zimet for helpful conversations.  We are grateful to Camillo De Lellis for discussions about geometric measure theory.  The work of M.D.~and L.M.~was supported in part by NSF grant PHY-1719877.  The work of C.L.~was supported in part by NSF grant PHY-1620526.
The work of M.S.~was supported in part by NSF grant DMS-1502294.  Portions of this work were completed at the Aspen Center for Physics, which is supported by National Science Foundation grant PHY-1607611.

\appendix

\section{Geometry of the Example}  \label{sec:app}
We consider the product $\tilde{V} = \mathbb{P}^1 \times \mathbb{P}^1 \times \mathbb{F}_2$. It can be realized as a toric variety obtained from the 4d reflexive polytope $\Delta^\circ$ with boundary points
\begin{equation}\label{eqn:mat1}
   \begin{pmatrix}
  \phantom{-}1 & -1 & \phantom{-}1 &-1 &-1 & -1 &-1 &\phantom{-}1 \\
  \phantom{-}1 &-1 & -1 & \phantom{-}1 &-1 & -1 & -1 &\phantom{-}1 \\
   -1 & \phantom{-}1 &  \phantom{-}1 &-1 & \phantom{-}0  & \phantom{-}0 &\phantom{-}0 &\phantom{-}0 \\
   \phantom{-}0 & \phantom{-}0 &  \phantom{-}1 & -1 & \phantom{-}2  &\phantom{-}1 & \phantom{-}0 &-1
  \end{pmatrix}
 \end{equation}
There is a unique triangulation $\mathcal{T}$ of the boundary of $\Delta^\circ$, which corresponds to a smooth toric fourfold $\widetilde{V}$. The maximal simplices of $\mathcal{T}$ are
\begin{eqnarray}\label{eq:cones}
&[[0, 2, 4, 5], [0, 2, 4, 7], [0, 2, 5, 6], [0, 2, 6, 7],\nonumber \\
&[0, 3, 4, 5], [0, 3, 4, 7], [0, 3, 5, 6], [0, 3, 6, 7],\nonumber \\
&[1, 2, 4, 5], [1, 2, 4, 7], [1, 2, 5, 6], [1, 2, 6, 7],\nonumber \\
&[1, 3, 4, 5], [1, 3, 4, 7], [1, 3, 5, 6], [1, 3, 6, 7]] \, .
\end{eqnarray}
The degrees in the Cox ring of $\widetilde{V}$ take the form

\begin{equation}\label{eqn:appcharges}
\begin{array}{|c|c|c|c|c|c|c|c|c|}
\hline
x_0 & x_1 & x_2 & x_3 & x_4 & x_5 & x_6 & x_7 & -K\\
\hline
1 & 1 &0 & 0 & 0 &0 & 0 & 0 & 2\\
0 & 0 &1 & 1 & 0 &0 & 0 & 0 & 2\\
0 & 0 &0 & 0 & 0 &1 & 0 & 1 & 2\\
0 & 0 &0 & 0 & 1 &-2 & 1 & 0 & 0\\
\hline
\end{array}\
\end{equation}
Here $x_i$ are the toric coordinates of $\widetilde{V}$, and $-K$ is the anticanonical class. The Stanley-Reisner ideal takes the form
\begin{equation}\label{eq:appsri}
\left< x_0 x_1, x_2 x_3, x_4 x_6, x_5 x_7 \right>\, .
\end{equation}
We consider an anticanonical hypersurface $\widetilde{X}$ in $\widetilde{V}$, defined by the vanishing of a polynomial $F$. A generic $F$ takes the form
\begin{equation}
F = x_5^2 P_{(2,2,0,4)} + x_5 x_7  P_{(2,2,0,2)} - x_7^2  P_{(2,2,0,0)}\, ,
\end{equation}
where the $P$'s are generic polynomials in the variables $(x_0, x_1,x_2,x_3,x_4,x_6)$. We will now construct an orientifold $X$ of this Calabi-Yau manifold by choosing a $\mathbb{Z}_2$ involution $\sigma$, such that $X \simeq \widetilde{X}/\sigma$. The general procedure is outlined in ~\cite{Collinucci:2008zs, Collinucci:2009uh}.
We choose an orientifold involution of the form $\sigma : x_5 \rightarrow -x_5$. To ensure the Calabi-Yau is invariant under this involution we require only even powers of $x_5$ to appear in $F$, so we take
\begin{equation}
F = x_5^2 P_{(2,2,0,4)} - x_7^2  P_{(2,2,0,0)}\,.
\end{equation}
In order to identify the fixed loci, we need to take into account the projective scalings of the coordinates in \eqref{eqn:appcharges}.
Under the action $x_5 \rightarrow - x_5$, from \eqref{eqn:appcharges} the fixed point loci are solutions of
\begin{align}
 x_0 &= \lambda_1 x_0 \nonumber \\
 x_1 &= \lambda_1 x_1 \nonumber \\
 x_2 &= \lambda_3 x_2 \nonumber \\
 x_3 &= \lambda_3 x_3 \nonumber \\
 x_4 &= \lambda_4 x_4 \nonumber \\
 x_5 &= -\lambda_2 \lambda_4^{-2} x_5 \nonumber \\
 x_6 &= \lambda_4 x_6 \nonumber \\
 x_7 &= \lambda_2 x_7\, ,
\end{align}
where $\lambda_i \in \mathbb{C}^{*}$. Via \eqref{eq:appsri} we see that $x_0$ and $x_1$ cannot vanish simultaneously, and so we can set $\lambda_1 = 1$. In addition, $x_2$ and $x_3$ cannot vanish simultaneously, and so $\lambda_3 = 1$. Finally,  $x_4$ and $x_6$ cannot vanish simultaneously, and so $\lambda_4 = 1$. We are then left with
\begin{align}
 x_5 &= -\lambda_2 x_5 \nonumber \\
 x_7 &= \lambda_2 x_7\, .
\end{align}
Since $x_5$ and $x_7$ cannot vanish simultaneously the solutions to this are $\lambda_2 = 1, x_5 = 0$ and $\lambda_2 = -1, x_7 = 0$. We will therefore have O7-planes on the subloci $x_5 = 0$ and $x_7 = 0$, but these O7-planes will not intersect. Following~\cite{Collinucci:2009uh}, we construct a new toric variety $V \simeq \widetilde{V}/\mathbb{Z}_2$ by writing down a basis of sections of the line bundles in \eqref{eqn:appcharges} that are invariant under $\sigma$, which will be the toric coordinates of $V$. This is achieved by taking
\begin{equation}
(x_0, x_1, x_2, x_3, x_4, x_5, x_6, x_7) \rightarrow (x_0, x_1, x_2, x_3, x_4, x_5^2, x_6, x_7^2)\, .
\end{equation}
We then define $h \equiv x_5^2$ and $\eta \equiv x_7^2$. Such a map is $2 \rightarrow 1$ everywhere away from the fixed loci, and $1 \rightarrow 1$ along the fixed loci. $V$ then has degrees in the Cox ring of the form
\begin{equation}\label{eqn:appcharges2}
\begin{array}{|c|c|c|c|c|c|c|c|c|}
\hline
x_0 & x_1 & x_2 & x_3 & x_4 & h & x_6 & \eta & F\\
\hline
1 & 1 &0 & 0 & 0 &0 & 0 & 0 & 2\\
0 & 0 &1 & 1 & 0 &0 & 0 & 0 & 2\\
0 & 0 &0 & 0 & 0 &1 & 0 & 1 & 1\\
0 & 0 &0 & 0 & 1 &-4 & 1 & 0 & 0\\
\hline
\end{array}\
\end{equation}
The Stanley-Reisner ideal takes the same form as before, with $x_2 \rightarrow h$, and $x_7 \rightarrow \eta$:
\begin{equation}\label{eq:appsri2}
\left< x_0 x_1, x_4 x_6, x_2 x_3, h \eta \right>\, .
\end{equation}
The new toric variety $V$ is also smooth, and has the same cone structure as given in \eqref{eq:cones}. From the degrees one can construct the rays of the corresponding fan,
\begin{equation}\label{eqn:mat2}
   \begin{pmatrix}
  \phantom{-}0 & \phantom{-}0 & \phantom{-}0 & \phantom{-}0 & \phantom{-}1 & \phantom{-}0 &-1 & \phantom{-}0  \\
  \phantom{-}0 &\phantom{-}0 & \phantom{-}1 &-1 & \phantom{-}0 & \phantom{-}0 & \phantom{-}0 & \phantom{-}0  \\
   \phantom{-}0 & \phantom{-}0 & \phantom{-}0 &\phantom{-}0 & \phantom{-}0  & \phantom{-}1& \phantom{-}4 & -1 \\
   \phantom{-}1 & -1 & \phantom{-} 0 & \phantom{-}0 &\phantom{-}0  &\phantom{-}0 &\phantom{-}0  &\phantom{-}0
  \end{pmatrix}
 \end{equation}
The orientifold $X \subset V$ is then defined by the vanishing of the polynomial $F$, with $x_5^2$ replaced by $h$ and $x_7^2$ replaced by $\eta$:
\begin{equation}\label{eqn:appnoncy2}
F = h P_{(2,2,0,4)} - \eta  P_{(2,2,0,0)} = 0\, .
\end{equation}
Note that $\{F = 0\}$ is not a Calabi-Yau hypersurface in $V$.

Let us briefly discuss the F-theory lift of this example.  The canonical bundle on $X$ can be computed from the adjunction formula:
\begin{equation}
K_X = (K_{V} + X)|_X\, .
\end{equation}
We have $\mathcal{O}_{\widetilde{V}}(K_{V} ) =\mathcal{O}_{V}(-2,-2,-2,2)$, and $\mathcal{O}_{V}(X )  =\mathcal{O}_{V}(2,2,1,0)$, and therefore $K_X \simeq \mathcal{O}_{V}(0,0,-1,2)|_X$.
The Weierstrass model is written as $y^2 = x^3 + fx + g$, with $f \in \Gamma(\mathcal{O}(-4K_X))$ and $g \in \Gamma(\mathcal{O}(-6K_X))$. In our example $f$ and $g$ both take particularly simple forms
\begin{align}
& f= h^2 \left( a_1 h^2 P_8 ( x_4, x_6) + a_2 h \eta P_4 ( x_4, x_6)  + a_3 \eta^2\right) , \nonumber \\
& g = h^3 \left(b_1 h^3 P_{12} ( x_4, x_6) + b_2 h^2 \eta  P^{\prime}_{8} ( x_4, x_6) + b_3 h \eta^2 P^{\prime}_{4} ( x_4, x_6) + b_4 \eta^3\right)\, ,
\end{align}
where the $a_i, b_i \in \mathbb{C}$. There is a non-Higgsable $I_0^{*}$ fiber on $h = 0$, consistent with the fact that $D_h$ is rigid.  In the particular case \begin{align}
& f= h^2  \eta^2 \nonumber \\
& g = h^3 \eta^3 \, ,
\end{align}
we have an $SO(8)$ gauge group on both $h = 0$ and $\eta = 0$, and no additional 7-branes in $X$. Such a case allows us to tune to arbitrarily
weak coupling globally on $X$, and so we expect the effective theory to be well-controlled.
The Euler characteristic can be computed by the method of~\cite{Esole:2017kyr}, and we find $\chi(Y) = 576$.
The D3-brane tadpole of this Calabi-Yau fourfold with gauge group $SO(8)^2$ is therefore $\chi(Y)/24 = 24$.

\section{Effective Divisors in the Example} \label{sec:effective}

In this appendix, we show that on  $X = \{F_{2,2,1,0} = 0\} \subset V = \PP^1 \times \PP^1 \times \FF_4$,
if $D = (-a,a,c,d)$ is an induced divisor class on $X$, for $a \in \ZZ_+, c, d \in \ZZ$,
then $D$ is not effective on $X$.
Recall that a divisor class $D$ on $X$ is called \emph{effective} if $H^0(X,
\cO_X(D)) \ne 0$.

It is more convenient, and hopefully makes the computations more clear, to generalize this
to a more natural setting. In fact, we show the following: suppose that $Y$ is any smooth toric
variety, that $\alpha \in \mathrm{Pic}(Y)$ is any effective divisor class, and that $F =
F_{2,2,\alpha}$ is a generic polynomial in the Cox ring $S$ of $V = \PP^1
\times \PP^1 \times Y$ of the given triple degree.  Let $X \subset V$
be the zero locus of $F$.  We assume that $X$ is smooth.
Given $(a,b,\gamma)$, where $a, b \in \ZZ$, and
$\gamma \in \mathrm{Pic}(Y)$, we obtain the pullback $\cO_X(a,b,\gamma)$ on $X$.

In this appendix, we show

\begin{theorem}~\label{thm:appendix}
  Fix an integer $a>0$.  For all $F$ in a dense Zariski-open subset of $S_{2,2,\alpha}$, and for
  all $\gamma \in \mathrm{Pic}(Y)$, the divisor class $(-a,a,\gamma)$ is {\em not} effective on $X$.
  \end{theorem}

We will first recall some basic facts about the cohomology of line bundles on toric varieties, then we analyze the relevant examples.

\subsection{Cohomology of line bundles on toric varieties}

We recall the description of the cohomology of line bundles on a toric variety.  For simplicity, we
now restrict to the case of simplicial toric varieties, but there is no problem generalizing
to arbitrary complete normal toric varieties.

Let $V \subset \PP_\Sigma$ be the simplicial toric variety
corresponding to a complete simplicial fan $\Sigma$,
with $r$ rays.  Let $D_1, \ldots, D_r$ denote the
corresponding divisors on $V$.
Suppose that $S = \C[x_1, \ldots,
  x_r]$ is the homogeneous coordinate ring of $V$ (Cox ring), and that
$I \subset S$ is the Stanley-Reisner ideal of $S$ corresponding to the
fan $\Sigma$.  The polynomial ring $S$ is graded by $\mathrm{Cl}(V)$, with
$\deg x_i = [D_i]$.

We often identify $\Sigma$ with a set of subsets
of $\rset := \{ 1, \ldots, r \}$.
Given a subset $\lambda \subset \rset$, define the simplicial complex $\Sigma_\lambda$ to
be
 \begin{equation} \Sigma_\lambda := \{ \tau \subset \Sigma \mid \tau \subset \lambda \}\,. \end{equation}

Consider the Laurent ring $S_{\mathbf x} := \C[\ZZ^r] = \C[x_1^{\pm
    1}, \ldots, x_r^{\pm 1}]$.  It turns out that the cohomology
groups of line bundles can often be identified with the $\C$-span of a
finite number of monomials in $S_{\mathbf x}$, or at least are closely related to
such sets.

For $u \in \ZZ^r$, let $\mathrm{neg}(u) := \{ i \in \rset \mid u_i < 0 \}$.
For $\lambda \subset \rset$ and $\alpha \in \mathrm{Cl}(V)$, define
\begin{equation} \cB(\lambda, \alpha) := \{ x^u \mid u \in \ZZ^r, \mathrm{neg}(u) = \lambda, \deg(u) = \alpha \}\,, \end{equation}
and denote the $\C$-vector space they generate by
\begin{equation} \mathrm{span}(\lambda, \alpha) := \mathrm{span}_\C \ \cB(\lambda, \alpha)\,. \end{equation}

The result we use is the following (see \cite{EMS2000, MS2004,cohomcalg}):
\begin{equation}\label{EMS2}
H^i(\cO_V(\alpha)) = \bigoplus_{\lambda \subset \rset}\mathrm{span}(\lambda, \alpha)
   \otimes \tilde{H}^{i-1}(\Sigma_\lambda, \C)\,.
\end{equation}
Let
\begin{equation} \label{Sigdef} \Sigma^i = \Sigma^i_V := \{ \lambda \subset \rset \mid \dim \widetilde{H}^{i-1}(\Sigma_\lambda) \ne 0 \}\,.\end{equation}
If $x^v \in S_\beta$ has degree $\beta \in \mathrm{Cl}(V)$, then $x^v$ defines a map of vector spaces
\begin{equation}\label{themap}
 H^i(\cO_V(\alpha-\beta)) \stackrel{x^v}{\longrightarrow} H^i(\cO_V(\alpha))\,.
\end{equation}
Suppose $\lambda \in \Sigma^i$ and $x^u \in \mathcal{B}(\lambda,\alpha-\beta)$.
Let $\mu = \mathrm{neg}(v+u)$.  Notice that $\mu \subset \lambda$, since all entries of $v$ are non-negative,
so $\Sigma_\mu \subset \Sigma_\lambda$,
and therefore there is a natural map
\begin{equation}\label{pidef}
\pi : \tilde{H}^{i-1}(\Sigma_\lambda) \longrightarrow \tilde{H}^{i-1}(\Sigma_\mu)\,.
\end{equation}
With the identification in \eqref{EMS2}, the map \eqref{themap} sends
\begin{equation} x^u \otimes e \in \mathrm{span}(\lambda, \alpha-\beta) \otimes \tilde{H}^{i-1}(\Sigma_\lambda) \end{equation}
to
\begin{equation} x^{u+v} \otimes \pi(e)\,. \end{equation}

In the examples that we consider in this appendix, it turns out that $\lambda \in \Sigma^1$ if and only if $\dim \widetilde{H}^0(\Sigma_\lambda) = 1$.  In this case,
a basis for $H^1(\cO_V(\alpha))$ is given by the set of monomials
\begin{equation} \cB(\alpha) := \bigcup_{\lambda \in \Sigma^1} \cB(\lambda, \alpha)\,.
\end{equation}

\subsection{The explicit example}

Recall the example of interest: let $Y$ be any smooth toric
variety, let $\alpha \in \mathrm{Pic}(Y)$ be any effective divisor class, and let $F =
F_{2,2,\alpha}$ be a generic polynomial in the Cox ring $S$ of $V = \PP^1
\times \PP^1 \times Y$ of the given triple degree.  Let $X \subset V$
be the zero locus of $F$, and assume that $X$ is smooth.

Since $\mathrm{Cl}(V) = \mathrm{Pic}(V) = \ZZ^2 \oplus \mathrm{Pic}(Y)$, we write degrees as triples
$(a, b, \gamma)$, where $a, b, \in \ZZ$ and $\gamma \in \mathrm{Pic}(Y)$.  The Cox ring of $V$ can be written as $S = R[x_1, x_2, x_3, x_4]$, where $R$ is the Cox ring
of $Y$, and degrees (over $S$) are: $\deg(x_1) = \deg(x_2) = (1,0,0)$,
$\deg(x_3) = \deg(x_4) = (0,1,0)$, and if $f \in R_\gamma$, then $\deg_S(f) = (0,0,\gamma)$.

Given $(a,b,\gamma) \in \mathrm{Pic}(V)$, where $a, b \in \ZZ$, and
$\gamma \in \mathrm{Pic}(Y)$, we obtain the pullback $\cO_X(a,b,\gamma)$ on $X$.
Using the long exact sequence in cohomology arising from the short exact sequence
\begin{equation}\begin{tikzcd}
0 \arrow[r] & \cO_V(a-2, b-2, \gamma - \alpha) \arrow[r, "F"] & \cO_V(a,b,\gamma) \arrow[r]
  & \cO_X(a,b,\gamma) \arrow[r] & 0\,,
  \end{tikzcd}
\end{equation}
it follows that
$(a, b, \gamma)$ is an effective divisor class on $X$ if and only if either it is
an effective divisor class on $V$, or the kernel of the map
\begin{equation} \phi_{a,b,\gamma} : H^1(\cO_V(a-2,b-2,\gamma-\alpha)) \longrightarrow
  H^1(\cO_V(a,b,\gamma)) \end{equation}
  is nonzero.

Our strategy for proving Theorem~\ref{thm:appendix} is the following.  We fix $a > 0$, and identify bases for the source and target $H^1$ modules for the map
$\phi_{-a,a,\gamma}$.  We sum over all $\gamma$, obtaining
a map $\phi_{-a,a,*}$ of graded $R$-modules, which turns out to be of
size $(a^2-1) \times (a^2-1)$, i.e.~a square matrix.  We show that $\mathrm{det}(\phi_{-a,a,*}) \neq 0$, which implies that for all $\gamma$ the kernel of $\phi_{-a,a,\gamma}$ is zero, and so
$(-a, a, \gamma)$ is not effective on $X$.  We
show that the determinant is nonzero by choosing a specialization of $F$, and bases of the source
and target, such that $\phi_{-a,a,*}$ is in block diagonal form, and each
block is (at worst) a tridiagonal matrix, whose determinant is nonzero.

\medskip\noindent
{\bf Proof of Theorem~\ref{thm:appendix}.}
Define
\begin{equation} H^1(\cO_V(a,b,*)) := \bigoplus_{\gamma \in \mathrm{Pic}(Y)} H^1(\cO_V(a,b,\gamma))\,. \end{equation}
  This is a finitely generated graded $R$-module.  Let $\phi_{a,b,*} := \bigoplus_{\gamma \in \mathrm{Pic}(Y)} \phi_{a,b,\gamma}$.
  Then
  \begin{equation} \phi_{a,b,*} : H^1(\cO_V(a-2, b-2, *)) \longrightarrow H^1(\cO_V(a,b,*)) \end{equation}
  is a graded $R$-module map, whose entries (once a basis is chosen) all lie in $R_\alpha$.

The set $\Sigma_V^1$ introduced in \eqref{Sigdef} is
\begin{equation} \Sigma_V^1 = \{ \{1,2\}, \{3,4\} \} \cup \Sigma^1_Y\,, \end{equation}
where $\Sigma^1_Y$ is a set of subsets only involving the indices of the variables in $R$.

Fixing $a > 0$, we will now consider degrees of the form $(-a, b, \gamma)$.
Given an integer $a >0$ and an integer $b$, define the set of (Laurent) monomials
  \begin{equation}  \cB_{a,b} := \left\{ \frac{x_3^k x_4^{\ell}}{x_1^{i} x_2^j} \mid k+\ell = b, i+j = a, k, \ell \ge 0, \ i, j \ge 1 \right\}. \end{equation}

\begin{lemma}
  For $a > 0$, $\cB_{a,b}$ is a (free) basis over $R$ of $H^1(\cO_V(-a,b,*))$.
  If $b < 0$, then $H^1(\cO_V(-a,b,*)) = 0$.
\end{lemma}

\medskip\noindent{\bf Proof.}
This follows from the discussion in the last subsection by the observation that if $d = (-a, b, \gamma)$, for $a > 0$, then
there is only one $\lambda \in \Sigma^1$ for which $\mathcal{B}(\lambda,d) \neq \emptyset$, namely
$\lambda = \{ 1, 2 \}$.
In this case, $\Sigma_\lambda$ consists of two disjoint points, and so $\widetilde{H}^0(\Sigma_\lambda, \C) = \C$.
A basis of $H^1(\cO_V(-a,b,\gamma))$ is given by the Laurent monomials in $\cB_{a,b}$, multiplied by any monomial in $R_\gamma$. $\square$

\medskip

Note that for $a > 0$ and $b \ge 0$, we have
$\dim_\C H^1(\cO_V(-a,b,\gamma)) = (a-1)(b+1) \dim R_\gamma$ and $\mathrm{rank}_R H^1(\cO_V(-a,b,*)) = (a-1)(b+1)$.
Also note that, for $a > 0$,
\begin{equation} \dim_\C H^1(\cO_V(-a-2,b-2,\gamma)) =
  \begin{cases}
    0 & \text{if $b \le 1$} \\
    (a+1)(b-1) \dim R_\gamma & \text{if $b \ge 1$}
  \end{cases}
\end{equation}

Computing the matrix of the map
  \begin{equation}\label{thenewmap} \phi_{-a,b,*} : H^1(\cO_V(-a-2, b-2, *)) \longrightarrow H^1(\cO_V(-a, b, *)) \end{equation}
induced by multiplication by the element $F \in S_{2,2,\alpha}$ with respect to these bases is
straightforward: if $x^u$ is a monomial in $S_{2,2,0}$, and $m$ is a monomial in $B_{-a-2,b-2}$, then
$m$ maps to either the monomial $x^u m$, if that is in $\cB_{-a,b}$, or 0, if not.  By linearity, we obtain
the matrix for the map \eqref{thenewmap}.

The following proposition is all that is needed to finish the proof of Theorem~\ref{thm:appendix}.

\begin{proposition}
  Fix $a > 0$.  Then $\phi_{-a, a, *}$ is a graded matrix of size $(a^2-1) \times (a^2-1)$ over
  the ring $R$, and for $F$ in a dense Zariski-open subset of $S_{2,2,\alpha}$, this determinant is nonzero.
  \end{proposition}

\noindent{\bf Proof.}
Notice that a basis (over $R$) for the source of the map $\phi_{-a,a,*}$ is $\cB_{a+2, a-2}$, a basis for the target is
$\cB_{a, a}$, and by inspection both of these have rank $a^2-1$.

Let $F \in S_{2,2,\alpha}$.  We will show that the
matrix $\phi_{-a,a,*}$ (with entries in $R$) has nonzero determinant,
for at least one $F$ of this degree, which implies that the
determinant is nonzero for a Zariski-open subset of $F \in
S_{2,2,\alpha}$, proving the result.

To do this, we choose a specific $F$ of the form
\begin{equation} F = x_1^2 x_3^2 A + x_1 x_2 x_3 x_4 B + x_2^2 x_4^2 C, \end{equation}
where $A, B, C \in R_\alpha$ are chosen generically (actually, if $x^v$ is any monomial in $R_\alpha$, then
we could choose $A$, $B$, $C$ to be $\C$-multiples of this monomial).

We now partition the bases of the source and target of $\phi_{-a,a,*}$ in a manner which will place
the resulting matrix in block form, with blocks that are square matrices whose determinants are evidently nonzero.  This is enough to prove the result.

To this end, let $t = \frac{x_2 x_4}{x_1 x_3}$, which allows us to write $F$ as:
\begin{equation} F = x_1 x_2 x_3 x_4 (t^{-1} A + B + t C)\,. \end{equation}

We define an equivalence relation on the set of monomials $m \in \mathcal{B}_{a,b}$
by setting $m \sim n$ if there is an $\ell \in \ZZ$ such that $n = t^\ell m$.
Define $\cB_{a,b,m} := \{ n \in \cB_{a,b} \mid n \sim m \}$.
Let $\widehat{\cB}_{a,b} := \{ m \in \cB_{a,b} \mid mt^{-1} \not\in \cB_{a,b} \}$.  It is easy to check that $\widehat{\cB}_{a,b}$ is a full set of representatives under this
equivalence relation, and that if $m \in \widehat{\cB}_{a,b}$, then
$\cB_{a,b,m} = \{ m, tm, t^2m, \ldots, t^im \}$, for some integer $i$
depending on $m$ (and also on $a$ and $b$).  It is also clear that
$\phi_{-a,b,*}$ maps the span of $\cB_{a+2,b+2,m}$ into the span of
$\cB_{a,b,x_1 x_2 x_3 x_4 m}$, and so becomes block diagonal with
these partitions of the bases.  The matrix
of $F$, on each $m$-set, has a tridiagonal structure.  When $a = b$,
in fact, one checks that each block is a square matrix, and is
(at most) tridiagonal, with determinant a nonzero polynomial in $A$, $B$, and $C$,
thus proving the proposition and hence the theorem. $\square$

\bibliographystyle{JHEP}
\bibliography{refs}

\end{document}